\documentclass[twocolumn,showpacs,preprintnumbers,prb]{revtex4}
\usepackage{amssymb}
\usepackage[dvips]{color}
\usepackage{epsfig}
\usepackage{dcolumn}
\usepackage{bm}

\def\G{\Gamma}
\def\d{\delta}
\def\D{\Delta}
\def\e{\epsilon}

\def\s{\sigma}

\def\wid{\widetilde}

\begin{document}

\title{Transient Charging and Discharging of Spin-polarized Electrons in a Quantum Dot}
\author{F. M. Souza,$^{1,2}$ S. A. Le{\~a}o,$^{3}$ R. M. Gester,$^{4}$ and A. P. Jauho$^{5,6}$}
\affiliation{$^1$ International Centre for Condensed Matter Physics, Universidade de Bras{\'i}lia, 70904-910, Bras{\'i}lia-DF, Brazil \\
$^2$ Centre for Advanced Study, Norwegian Academy of Science and
Letters, Drammensveien 78, NO-0271 Oslo, Norway \\ $^3$ Instituto
de F\'{\i}sica,
Universidade Federal de Goi{\'a}s, 74001-970, Goi{\^a}nia-GO, Brazil \\
$^4$Grupo de F{\'i}sica de Materiais da Amaz{\^o}nia, Departamento
de F{\'i}sica, Universidade Federal do Par{\'a}, 66075-110,
Bel{\'e}m-PA, Brazil \\ $^5$ MIC - Department of Micro and
Nanotechnology, NanoDTU, Technical University of Denmark, {\O}rsteds
Plads, Bldg. 345E, DK-2800 Kgs. Lyngby, Denmark \\ $^6$ Laboratory
of Physics, Helsinki University of Technology, P. O. Box 1100,
FI-02015 HUT, Finland}  \pacs{PACS number}

\begin{abstract}
We study spin-polarized transient transport in a quantum dot coupled
to two ferromagnetic leads subjected to a rectangular bias voltage
pulse. Time-dependent spin-resolved currents, occupations, spin
accumulation, and tunneling magnetoresistance (TMR) are calculated
using both nonequilibrium Green function and master equation
techniques. Both parallel and antiparallel leads' magnetization
alignments are analyzed. Our main findings are: a dynamical spin
accumulation that changes sign in time, a short-lived pulse of spin
polarized current in the emitter lead (but not in the collector
lead), and a dynamical TMR that develops negative values in the
transient regime. We also observe that the intra-dot Coulomb
interaction can enhance even further the negative values of the TMR.
\end{abstract}

\volumeyear{year} \volumenumber{number} \issuenumber{number}
\eid{identifier}
\date[Date: ]{\today}
\maketitle

\section{Introduction}

A variety of new effects and novel devices have been reported
during recent years in the context of the emerging field of
spintronics.\cite{spintronics,iz04,saw01,gap98} One of the most
challenging milestones in this context is the development of a
quantum computer, which would represent a great breakthrough in
the processing time of certain mathematical and physical
problems.\cite{quantumcomp} In particular, the electron spin in
quantum dots has been proposed as a building block for the
implementation of quantum bits (qubits) for quantum
computation.\cite{dl98,hae05} An important recent development is
the possibility to coherently control electron states and electron
spin in quantum dot systems with a precision up to a
single-electron, thus demonstrating the feasibility of qubit
implementation in a solid state
system.\cite{th03,jme04,mk04,rh05,fhlk06,lpk06} Specifically,
these experimental realizations use high-speed voltage pulses to
tune the system levels in a coherent cycle for electronic
manipulation. Ac-driven quantum dot systems and double barrier
structures have also been studied in the context of quantum
pumps,\cite{js07,es06,ec05,la05,ms99} superlattices,\cite{rl03}
Kondo effect,\cite{rl98,k200,rl01,yy01} and spin-polarized
transport.\cite{zgz04,cl04} In addition to this, time-dependent
transport has received growing attention in a variety of
mesoscopic systems that encompasses, to mention but a few,
molecular electronics,\cite{cck05,khk06} dissipative driven
mesoscopic ring,\cite{la04} noisy qubits,\cite{fkh05} and
dynamical Franz-Keldysh effect.\cite{apj96}

In the context of spintronics a system of particular interest is
composed of a quantum dot or a metallic island coupled via tunnel
barriers to two ferromagnetic leads (FM-QD-FM). For example, in
the nonequilibrium regime the following effects have been
discussed: a spin-split Kondo resonance,\cite{yu05,jm03} a
spin-current diode effect,\cite{fms06} zero-bias
anomaly,\cite{iw05} tunnel magnetoresistance (TMR)
oscillations,\cite{iw06jb98} negative TMR, \cite{jv05} spin
accumulation,\cite{spinacum} and so on. In spite of all this
activity, to the best of our knowledge, only very little work has
been done on spin-polarized transport driven by ac-bias
voltages.\cite{zgz04,cl04} Here we study transient spin-resolved
currents, occupations and TMR generated by a voltage pulse applied
in one of the ferromagnetic leads. We use two complementary
approaches to study the problem: nonequilibrium Green function
(NEGF) and the Master Equation (ME). NEGF is used to give an exact
solution in the noninteracting case, while the ME, valid in the
limit $k_B T \gg \G_0$ ($\G_0$ is the characteristic level width),
is used to demonstrate that the results obtained via NEGF are
modified only quantitatively, not qualitatively, when Coulomb
interaction is accounted in the sequential-tunneling limit. Both
parallel (P) and antiparallel (AP) magnetization alignments are
considered. In the P case we find a magnitude and sign modulation
of the spin accumulation in the dot, while in the AP alignment
only the magnitude changes. For the current we observe a spike of
spin polarized current in the emitter lead when the system
operates in the P configuration. This effect gives rise to a
\emph{dynamical negative-TMR} just after the bias voltage is
turned off.

The paper is organized as follow. In Sec. II we describe the
formulation based on NEGF, and give explicit formulas for the
noninteracting case. In Sec. III(a)-(c) we present numerical
results based on Sec. II, and in Sec. III(d) we apply the master
equation technique to account Coulomb interaction effects (in the
sequential tunneling limit). Finally, in Sec. IV we give some
final remarks.

\begin{figure}[h] 
\par
\begin{center}
\epsfig{file=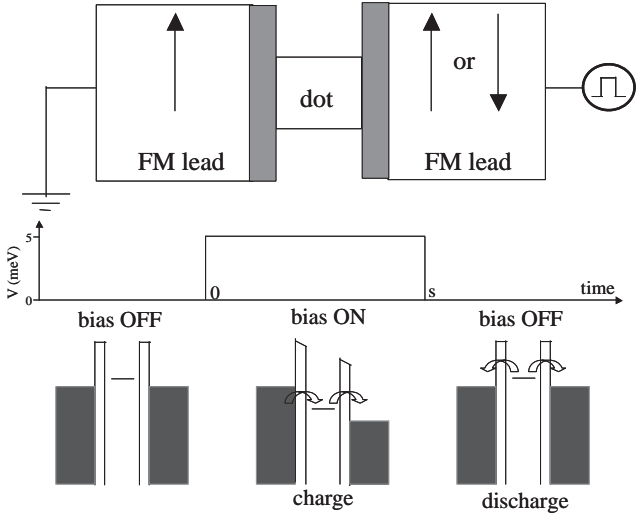, width=0.47\textwidth}
\end{center}
\caption{Sketch of the system: a quantum dot coupled to two
ferromagnetic leads via tunnel barriers. The left FM lead has its
magnetization fixed while the right-hand side can be either in
parallel or antiparallel alignment. A pulsed bias voltage of
duration $s$ is applied across the system in order to generate
transient spin-polarized currents. When the bias voltage is turned
on ($0<t<s$) the dot's level $\e_d$ moves into resonance with the
emitter states, and the dot becomes populated (a charging process)
with a current passing through it. When the bias is turned off
($t>s$) $\e_d$ moves above $\mu_L$ and $\mu_R$ and the dot's
occupation decays into the leads (a discharging process). Due to
the ferromagnetism of the leads these transient charging and
discharging processes become spin dependent.} \label{fig1}
\end{figure}

\section{Transport formulation}

To describe the system of a quantum dot coupled to two
ferromagnetic leads, see Fig. \ref{fig1}, we apply the following
Hamiltonian
\begin{eqnarray}
H&=&\sum_{\mathbf{k} \s \eta} \e_{\mathbf{k} \s \eta}(t)
c_{\mathbf{k} \s \eta}^\dagger c_{\mathbf{k} \s \eta}+\sum_{\s}
\e_d(t) d_\s^\dagger d_\s\nonumber\\
&\quad&+\sum_{\mathbf{k} \s \eta} (V_{\mathbf{k} \s \eta,\s}
c_{\mathbf{k}\s \eta}^\dagger d_\s +V_{\mathbf{k} \s \eta,\s}^*
d_\s^\dagger c_{\mathbf{k} \s
\eta})\nonumber\\
&\quad& +Un_{\uparrow}n_{\downarrow},\label{hamilt}
\end{eqnarray}
where $\e_{\mathbf{k} \s \eta}(t)$ is a time-dependent
free-electron energy with wave vector $\mathbf{k}$ and spin $\s$
in lead $\eta$ ($\eta=L, R$). This energy can also be written as
$\e_{\mathbf{k} \s \eta}(t)=\e_{\mathbf{k} \s \eta}^0+ \Delta_\eta
(t)$, with $\e_{\mathbf{k} \s \eta}^0$ being the time-independent
energy and $\Delta_\eta (t)$ gives the time evolution of the
external bias. The energy $\e_d (t)$ is the time-dependent
spin-degenerate dot level, which can also be written as $\e_d
(t)=\e_d^0 + \Delta_d (t)$, where $\e_d^0$ is the time-independent
level and $\Delta_d (t)$ follows the bias voltage.  It should be
noted that in a quantitative theory one should consider a
level-shift $\Delta_d$, which depends on the level occupation, via
some suitable self-consistent procedure.  We shall address this
issue in our future work, but for the present purpose the simple
model suffices.

The operator $c_{\mathbf{k} \s \eta}$ ($c_{\mathbf{k} \s
\eta}^\dagger$) is an annihilation (creation) operator for a
single-particle momentum state $\mathbf{k}$ and spin $\s$ in lead
$\eta$ ($\eta=L,R$), and $d_\s$ ($d_\s^\dagger$) is an
annihilation (creation) operator for the single-particle dot's
state $\e_d$. The matrix element $V_{\mathbf{k} \s \eta,\s}$
couples the leads with the dot, and we assume that the tunneling
process is spin-independent. Finally, the $U$-term describes the
Coulomb repulsion in the dot, with $n_\s=d_\s^\dagger d_\s$.

In order to calculate the current we use the definition
$I_\s^\eta=-e \langle \dot{N}_\s^\eta \rangle$, where $e$ is the
electron charge ($e>0$) and $N_\s^\eta= \sum_{\mathbf{k}}
c_{\mathbf{k} \s \eta}^\dagger c_{\mathbf{k} \s \eta}$ is the total
number of electrons with spin $\s$ in lead $\eta$. From this
definition it is straightforward to show that\cite{apj94,hh96}
\begin{equation}\label{Iseta}
I_\s^\eta(t)=2e \mathrm{Re} \{ \sum_{\mathbf{k}} V_{\mathbf{k} \s
\eta,\s} G_{\s,\mathbf{k}\s\eta}^<(t,t)\},
\end{equation}
where
\begin{eqnarray}\label{Glesser}
G_{\s,\mathbf{k}\s\eta}^<(t,t)&=&i \int_{-\infty}^{t} dt_1
V_{\mathbf{k} \s \eta,\s}^* e^{-i \int_{t}^{t_1}dt_2 \e_{\mathbf{k}
\s \eta}(t_2) }\nonumber\\&\quad&\times[G_{\s\s}^r(t,t_1) f_\eta
(\e_{\mathbf{k} \s}^0)+G_{\s\s}^<(t,t_1)],
\end{eqnarray}
with $G_{\s\s}^{r(<)}(t,t_1)$ being the retarded (lesser) Green
function of the dot and $f_\eta (\e_{\mathbf{k} \s}^0)$ is the
time-independent Fermi distribution function of lead $\eta$.
Substituting Eq. (\ref{Glesser}) into Eq. (\ref{Iseta}) and
following Ref. [\onlinecite{apj94}] we find
\begin{eqnarray}\label{Iseta2}
I_\s^\eta (t)&=& - 2e \int_{-\infty}^{t} dt_1 \int \frac{d\e}{2\pi}
\mathrm{Im} \{ e^{i \e (t-t_1)} \G_\s^\eta (\e,t_1,t)
\nonumber\\&\quad&\times[G_{\s\s}^r(t,t_1) f_\eta (\e) + G_{\s\s}^<
(t,t_1)]\},
\end{eqnarray}
with $\G_\s^\eta (\e,t_1,t)= 2 \pi \rho_{\s \eta} (\e) |V_{\s
\eta}(\e)|^2 e^{i \int_{t_1}^{t}dt_2 \D_\eta (\e,t_2)}$. These
results are exact, and they can in principle be used to study the
intricate interplay between time-dependence, coherence and
interactions. Their use, however, requires the knowledge of $G^r$
and $G^<$, which come from the solution of the nonequilibrium
Dyson and Keldysh equations, respectively. For our main findings,
though, it is sufficient to consider a non-interacting model, for
which an exact solution can be obtained. Next, in Sec. 3D, we show
that our results change only slightly when Coulomb interaction is
included in a master equation based scheme.

In the wideband limit (WBL),\cite{wbl,maciejko06} and for
noninteracting electrons Eq. (\ref{Iseta2}) can be written as
\begin{eqnarray}\label{Iseta3}
I_\s^\eta(t)=- e \G_\s^\eta \{ \langle n_\s (t) \rangle + \int
\frac{d\e}{\pi} f_\eta (\e) \mathrm{Im} [A_{\s \eta}(\e,t)]\},
\end{eqnarray}
where $\langle n_\s \rangle$ is the time-dependent dot's
occupation, given by
\begin{eqnarray}\label{nst}
\langle n_\s(t) \rangle &=& \mathrm{Im} \{ G_{\s\s}^< (t,t)
\}\nonumber\\&=&\sum_\eta \G_\s^\eta \int \frac{d\e}{2\pi} f_\eta
(\e) |A_{\s \eta} (\e,t)|^2,
\end{eqnarray}
and $A_{\s \eta}(\e,t)$ is defined as
\begin{equation}
A_{\s \eta}(\e,t)=\int_{-\infty}^{t} dt_1 G_{\s\s}^r(t,t_1) e^{[i \e
(t-t_1) - i \int_{t}^{t_1} d\wid{t} \D_\eta (\wid{t})]}.
\end{equation}
The retarded Green function in the noninteracting model is given
by
\begin{equation}
G_{\s\s}^r(t,t_1)=-i \theta(t-t_1) e^{-\frac{\G_\s}{2}(t-t_1)}
e^{-i \int_{t_1}^{t} d\wid{t} \e_d(\wid{t})},
\end{equation}
where $\G_\s=\G_\s^L+\G_\s^R$. For a voltage pulse $V(t)=V_0
\theta(t) \theta (s-t)$ (see Fig. \ref{fig1}), and assuming that
this pulse is applied on the right ferromagnetic lead, with a
linear bias drop along the junction, we have $\D_L(t) = -V_L=0$,
$\D_R(t) = -V_R(t)=-V(t)$ and $\D_d=-V_d=-V(t)/2$. 

With these definitions, we
find for $0<t<s$\cite{nsw93}
\begin{eqnarray}\label{A1}
A_{\s \eta} (\e,0<t<s)&=&
\frac{e^{i(\e-\e_0+V_d-V_\eta+i\G_\s/2)t}}{\e-\e_0+i
\G_\s/2}\nonumber\\&+&
\frac{1-e^{i(\e-\e_0+V_d-V_\eta+i\G_\s/2)t}}{\e-\e_0+V_d-V_\eta+i\G_\s/2},
\end{eqnarray}
and for $t>s$ we obtain
\begin{eqnarray}\label{A2}
&&A_{\s \eta} (\e,
t>s)=\frac{e^{i(\e-\e_0+i\G_\s/2)t}e^{i(V_d-V_\eta)s}}{\e-\e_0+i
\G_\s/2}\nonumber\\&+&\frac{e^{i(\e-\e_0+i\G_\s/2)(t-s)}-e^{i(V_d-V_\eta)s}
e^{i(\e-\e_0+i\G_\s/2)t}}{\e-\e_0+V_d-V_\eta+i
\G_\s/2}\nonumber\\&+&\frac{1-e^{i(\e-\e_0+i\G_\s/2)(t-s)}}{\e-\e_0+i
\G_\s/2}.
\end{eqnarray}
By substituting Eqs. (\ref{A1}) and (\ref{A2}) into Eqs.
(\ref{Iseta3})-(\ref{nst}) yields the final result for the
spin-resolved occupations and currents. Numerical results are
described in the next section.

\section{Results}

\subsection{Parameters}

In our numerical calculations we assume that the voltage pulse is
applied to the right electrode, so that $\mu_R=-V(t)$ while
$\mu_L$ is kept constant equal zero. The dot's level is taken
originally (zero bias) above the chemical potentials $\mu_L$ and
$\mu_R$, $\e_0=0.5$ meV. The temperature is assumed to be $T=2.5K$
$(k_B T \approx 215 \mu$eV), thus allowing a small thermally excited
occupation of the dot in equilibrium. To describe the
ferromagnetism of the leads we choose the tunneling rates to be
$\G_\s^L = \G_0 [1 + (-1)^{\d_{\downarrow \s}} p]$ and $\G_\s^R =
\G_0 [1 \pm (-1)^{\d_{\downarrow \s}} p]$, where $\G_0$ is the
leads-dot coupling strength and $p$ gives the polarization degree
of the leads.\cite{wr01} Here we assume a weak coupling with
$\G_0=1$ $\mu$eV,\cite{commentGamma,typical} and a polarization
degree $p=0.4$. The $+$ and $-$ signs in $\G_\s^R$ give the
parallel and antiparallel configurations, respectively. Due to the
ferromagnetism of the leads ($p \neq 0$) we have $\G_\uparrow^L >
\G_\downarrow^L$ and $\G_\uparrow^R > \G_\downarrow^R$ in the
parallel case and the opposite $\G_\uparrow^R < \G_\downarrow^R$
in the antiparallel alignment. For the bias voltage we adopt
$V(t)=V_0 \theta(t) \theta(s-t)$ where $V_0=5$ meV and $s=3$
ns.\cite{kbk04} The charging energy $U$ is set equal to zero in
Sec. III(b)-(c) and equal to 3 meV in Sec. III(d).

\begin{figure}[tbp]
\par
\begin{center}
\epsfig{file=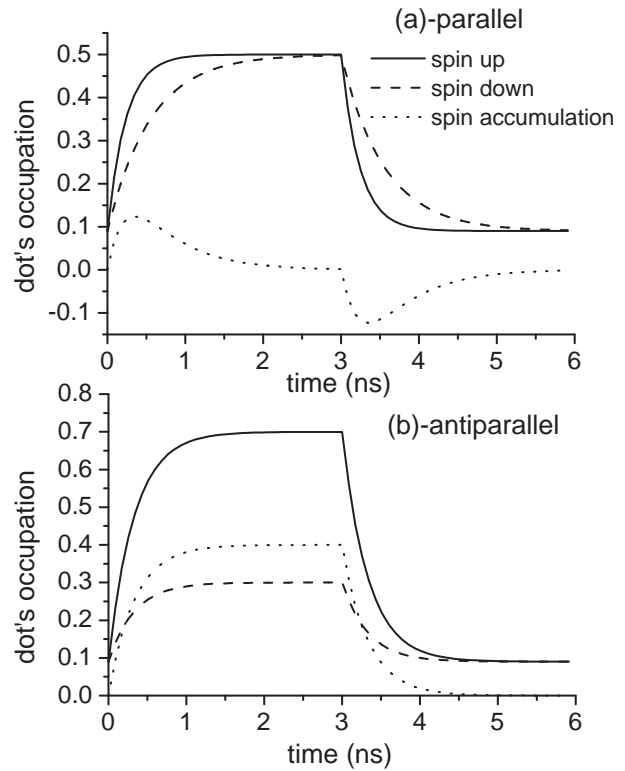, width=0.45\textwidth}
\end{center}
\caption{Occupations $n_\uparrow$ (solid line) and $n_\downarrow$
(dashed line) and the spin accumulation $m=n_\uparrow -
n_\downarrow$ (dotted line) as a function of time for both (a)
parallel and (b) antiparallel configurations. When the bias is
turned on (off) the dot is charged (discharged) in a
spin-dependent manner. This results in a time-dependent spin
accumulation, with a sign reversal in the parallel case.}
\label{fig2}
\end{figure}

\subsection{Spin-polarized occupations}

Figure \ref{fig2} shows the spin-resolved occupations $n_\uparrow$
and $n_\downarrow$ and the spin accumulation $m=n_\uparrow -
n_\downarrow$ as a function of time for both (a) parallel and (b)
antiparallel configurations. Before the bias is turned on the level
$\e_d$ is above the electrochemical potentials $\mu_\eta$
($\eta=L,R$), and the dot is only slightly occupied due to thermal
excitation. When the bias is turned on at $t=0$ the dot's level is
brought into resonance ($\mu_L < \e_d < \mu_R$), thus resulting in
an enhancement of $n_\s$ and $m$. In the parallel case [Fig.
\ref{fig2}(a)] the spin up population increases faster than the spin
down one, and both attain the same stationary value around 0.5. The
steeper enhancement of $n_\uparrow$ compared to $n_\downarrow$ is
related to the inequality $\G_\uparrow^L > \G_\downarrow^L$, that
gives a faster response for the spin $\uparrow$ component. Since
$\G_\s^L = \G_\s^R$ in the P case, the in- and out-tunnel rates
compensate each other, thus resulting in $n_\uparrow = n_\downarrow$
for asymptotic times. When the bias voltage is turned off, $\e_d$
raises above $\mu_L$ and $\mu_R$ and the population of the dot
begins to decay, with a faster discharge for the $\uparrow$
component. The spin accumulation reflects the dynamics of
$n_\uparrow$ and $n_\downarrow$. In the range $0<t<s$, $m$ reaches a
local maximum due to the faster enhancement of $n_\uparrow$ compared
to $n_\downarrow$. In contrast, when the bias voltage is turned off
($t>s$), $m$ shows a local (negative) minimum due to the fast
discharge of $n_\uparrow$.

\begin{figure}[h]
\par
\begin{center}
\epsfig{file=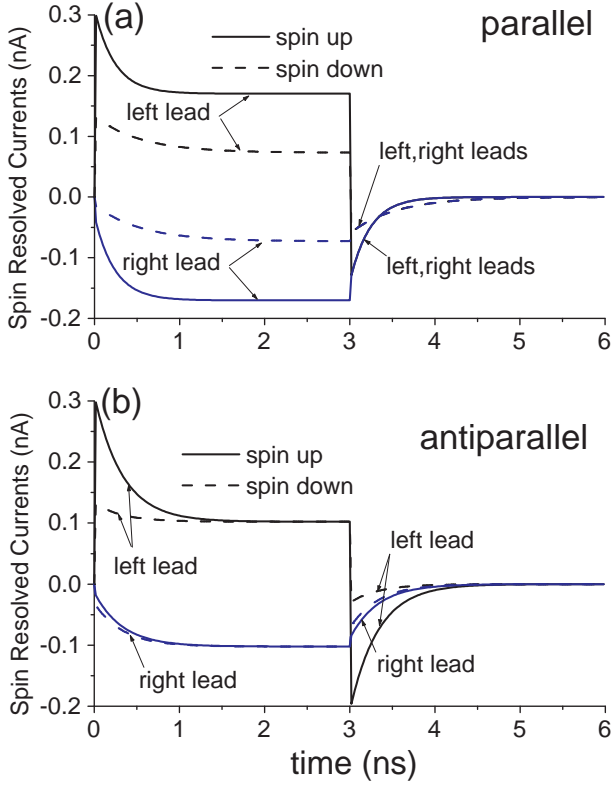, width=0.45\textwidth}
\end{center}
\caption{Spin resolved currents against time for both left and right
leads and both alignments. In both configurations the currents in
the left (right) lead are suppressed (enhanced) just after the
voltage is turned on, and then they attain stationary plateaus. When
the bias voltage is turned off ($t>3$ns) the left and right currents
become the same for each spin component in the P configuration,
while in the AP case the $\uparrow$ current becomes bigger than the
$\downarrow$ current in the left lead.} \label{fig3}
\end{figure}

In Figure \ref{fig2}(b) we show the evolution of the occupations and
the spin accumulation in the antiparallel alignment. We note that
$n_\uparrow$ increases faster than $n_\downarrow$ as in the P case.
In contrast, though, $n_\uparrow$ attains a higher value than
$n_\downarrow$ in the stationary regime. This is related to the
out-tunnel rates which are now inverted with respect to the parallel
case: $\G_\uparrow^R < \G_\downarrow^R$. When the bias is turned off
both $n_\uparrow$ and $n_\downarrow$ decrease due to the transient
discharge. In particular the spin up electron population discharges
predominantly to the left lead while the spin down component
discharges to the right, following their corresponding majority
density of states (or equivalently the majority tunnel rates). The
way how spins $\uparrow$ and $\downarrow$ charge and discharge are
more clearly seen in the spin-resolved current curves described in
the next section.

\subsection{Spin-resolved currents}

Figure \ref{fig3} shows $I_\uparrow$ and $I_\downarrow$ for both
leads and both ferromagnetic alignments. In the P configuration
[Fig. \ref{fig3}(a)] the left currents $I_\uparrow^L$ and
$I_\downarrow^L$ show a transient suppression and then attain
their respective stationary values with $I_\uparrow^L >
I_\downarrow^L$. In the right lead the currents $I_\uparrow^R$ and
$I_\downarrow^R$ increase (in modulus) up to their respective
stationary values. When the bias voltage is turned off $I_\s^L$
becomes negative as $I_\s^R$. The negative sign of both $I_\s^L$
and $I_\s^R$ means that the electrons are flowing from the dot to
the leads (discharge). In particular the spin $\downarrow$
electrons discharge much slower than the $\uparrow$ ones, due to
$\G_\downarrow^{L,R} < \G_\uparrow^{L,R}$.

\begin{figure}[tbp]
\par
\begin{center}
\epsfig{file=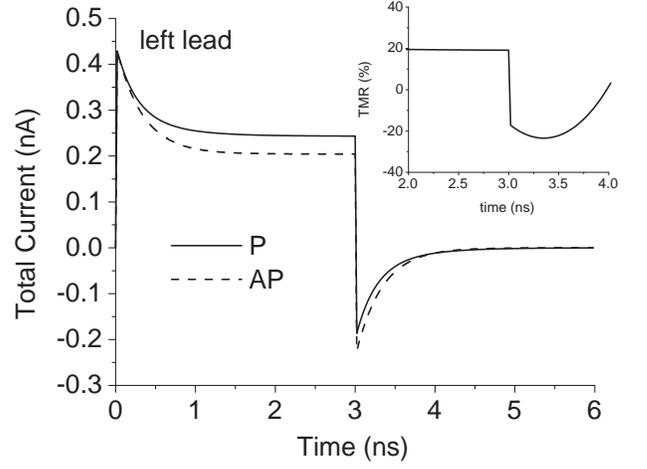, width=0.45\textwidth}
\end{center}
\caption{Total current in the left ferromagnetic lead,
$I_\uparrow^L+I_\downarrow^L$, for both parallel (solid line) and
antiparallel (dotted line) configurations. After the bias voltage
is turned off ($t>3$ ns) the total antiparallel current becomes
greater than the parallel one, lifted by the spike of $\uparrow$
current seen in Fig. \ref{fig3}(b). This results in the
time-dependent negative TMR seen in the inset.} \label{fig4}
\end{figure}

In the AP configuration [Fig. \ref{fig3}(b)] $I_\uparrow^L$ and
$I_\downarrow^L$ show a suppression just after the bias voltage is
turned on, then they attain a stationary value with $I_\uparrow^L
= I_\downarrow^L$. In the right lead the currents $I_\uparrow^R$
and $I_\downarrow^R$ are enhanced until they reach equal plateaus.
When the bias voltage is turned off, $I_\uparrow^L$ and
$I_\downarrow^L$ change sign (discharge of the dot) and a
\emph{spike} of spin $\uparrow$ current in seen in the left lead
($I^L_\uparrow \gg I^L_\downarrow$). This reflects the
preferential discharge of spin up electrons to the left lead,
according to $\G_\uparrow^L > \G_\uparrow^R$. No spike is seen in
the parallel configuration, where spin up electrons discharge
equally to both leads. In contrast, in the AP alignment the spin
down electrons discharge preferentially to the right lead due to
the inverted inequality $\G_\downarrow^L < \G_\downarrow^R$, while
in the P case its discharge is equally to both sides
($\G_\downarrow^L = \G_\downarrow^R$).

\emph{Negative TMR}. In figure (\ref{fig4}) we show the total
current in the left lead ($I_\uparrow^L+I_\downarrow^L$) for both
parallel and antiparallel configurations. Due to the strong
spin-polarized discharge ($t>3$ ns) in the left lead when the
system is AP aligned, the total current obeys the unusual
inequality $I_{AP}^L > I_{P}^L$, which results in
\emph{time-dependent negative tunnel magnetoresistance} (see
inset), defined as $TMR=(I_P^L-I_{AP}^L)/I_{AP}^L$. As the time
evolves the TMR keeps increasing, due to the longer spin-down
lifetimes when the system is parallel aligned. More specifically,
in the AP configuration both spin up and down discharge fast to
the left and to the right leads, respectively, following their
majority spin populations (or equivalently the tunneling rates).
In contrast, in the P alignment the majority populations occur for
spin up in both leads ($\G_\uparrow^{L,R} > \G_\downarrow^{L,R}$).
This turns into a fast discharge for spin up electrons and a slow
discharge for the down component. This slow spin down discharge
sustains the total current much longer than in the AP
configuration, and eventually for long enough times we find $I_P^L
\gg I_{AP}^L$.

\emph{Displacement Current}. In the transient regime the left and
the right currents are not in general the same ($I_L \neq I_R$),
due to charge accumulation/depletion in the dot. The generalized
conservation law is given by the continuity equation
$I^L_\s+I^R_\s-I^{dis}_\s=0$, where $I^{dis}_\s$ is the
displacement current for spin $\s$, given by $I^{dis}_\s=ed
\langle n_\s(t) \rangle /dt$. In order to check the accuracy of
our numerical calculation we have verified numerically the
continuity equation.

\subsection{Effects of Coulomb Interaction}

An exact treatment of the Coulomb interaction represents a
formidable problem, and in the context of the present Hamiltonian
only few results are known \emph{in equilibrium}, and none in
nonequilibrium, even less so under transient conditions.
Nevertheless, in certain limits approximate treatments may give a
good qualitative understanding of the generic behavior.  One such
case is the sequential-tunneling limit ($\G_0 \ll k_B T$), where
the Master Equation (ME) approach is known to work well. Here, we
use the ME to estimate the effects of Coulomb interaction in our
results.\cite{timedepenU} The current expression is given
by\cite{mastereq}
\begin{equation}\label{ImasterEq}
I_\s^\eta = e \G_\s^\eta [f_\eta P_0 - (1-f_\eta) P_\s +
\tilde{f}_\eta P_{\bar{\s}} - (1-\tilde{f}_\eta) P_2],
\end{equation}
where $P_0=\langle (1-n_\uparrow) (1-n_\downarrow) \rangle$,
$P_\s=\langle n_\s (1-n_{\bar{\s}})\rangle$ and $P_2=\langle
n_\uparrow n_\downarrow \rangle$, are the probabilities to have no
electron, one electron with spin $\s$ and two electrons,
respectively. The Fermi functions $f_\eta$ and $\tilde{f}_\eta$
are evaluated at $\e_d$ and $\e_d+U$, respectively. For the dot's
occupation we write
\begin{eqnarray}
\frac{d}{dt} \langle n_\s \rangle &=&
\frac{1}{e}(I_\s^L+I_\s^R)\nonumber\\&=&\sum_{\eta}\G_\s^\eta
[f_\eta P_0 - (1-f_\eta) P_\s \nonumber\\
&\quad&+ \tilde{f}_\eta P_{\bar{\s}} - (1-\tilde{f}_\eta)
P_2],
\end{eqnarray}
and for the double occupancy probability we have
\begin{equation}
\frac{d}{dt}\langle n_\uparrow n_\downarrow \rangle = \sum_{\s
\eta} \G_{\s}^\eta [\tilde{f}_\eta P_{\bar{\s}}-(1-\tilde{f}_\eta)
P_2].
\end{equation}

\begin{figure}[tbp]
\par
\begin{center}
\epsfig{file=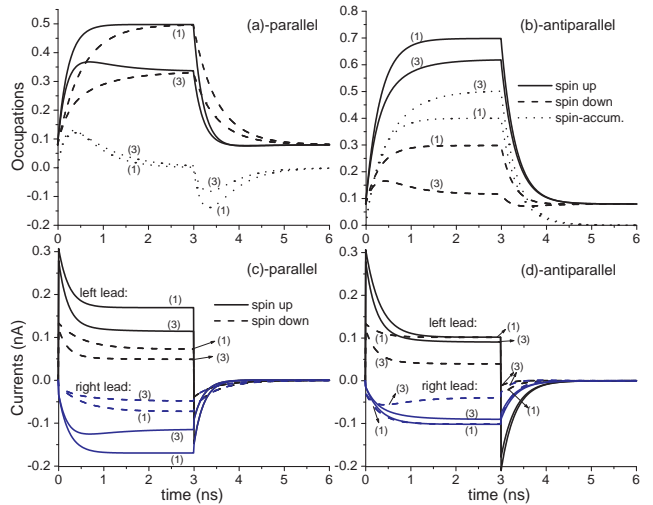, width=0.47\textwidth}
\end{center}
\caption{Spin-resolved occupations (a)-(b) and currents (c)-(d) in
both parallel (P) and antiparallel (AP) configurations. We take
$U=1$ meV and $U=3$ meV, labeling the traces by (1) and (3),
respectively. For $U=1$ meV the results are indistinguishable from
the $U=0$ case. For $U=3$ meV, though, we find a suppression of
the spin-resolved occupations and currents. In the AP alignment
this suppression results in an enhancement of the spin
accumulation and in a spin polarized current in the stationary
plateaus ($|I_\uparrow^{L,R}|>|I_\downarrow^{L,R}|$). To clarify
the range $t>3$ ns, in panel (c) the currents $I_\s^L$ are on top
of $I_\s^R$ while in panel (d) they are apart from each other. In
addition in the AP case, $|I_\uparrow^{L}|$ and $|I_\uparrow^{R}|$
for $U=1$ meV are slightly greater than their corresponding values
for $U=3$ meV, and $I_\downarrow^R$ is almost on top of
$I_\uparrow^R$.} \label{fig5}
\end{figure}

For the noninteracting case ($U=0$) the time-dependent results
obtained from Eq. (\ref{ImasterEq}) are identical to those seen in
Sec. III(b)-(c). For the interacting case ($U \neq 0$), we find
that for $U=1$ meV the results are indistinguishable from the
$U=0$ case [see Fig. \ref{fig5}]. This is so because for small
enough $U$ both channels $\e_d$ and $\e_d+U$ attain resonance for
$V(t) = 5$ meV. In contrast, for $U=3$ meV the channel $\e_d+U$
remains above the emitter chemical potential when the bias voltage
is applied, which turns into a suppression of the occupations and
the currents. In particular in the AP configuration this
suppression is stronger upon the spin down component, seen in both
occupations [panel (b)] and currents [panel (d)]. This is due to
the spin imbalance $n_\uparrow
> n_\downarrow$ typically present in the antiparallel alignment.
This spin-polarized suppression in the AP configuration gives rise
to an enhancement of the spin imbalance [see Fig. \ref{fig5}(b)]
and to a spin polarized current
$(|I_\uparrow^{L,R}|>|I_{\downarrow}^{L,R}|)$ in the stationary
plateau.

\begin{figure}[tbp]
\par
\begin{center}
\epsfig{file=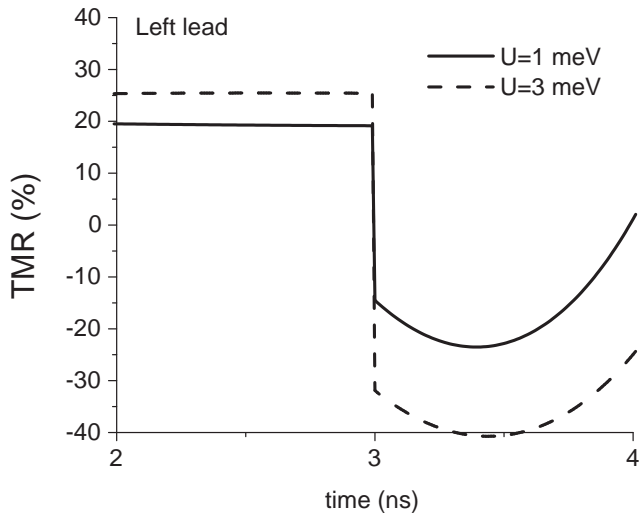, width=0.47\textwidth}
\end{center}
\caption{Tunnel magnetoresistance (TMR) as a function of time for
$U=1$ and $U=3$ meV. For $U=3$ meV the TMR is enhanced in the
stationary regime (2$<t<$3 ns) and becomes even more negative
after the bias voltage is turned off ($t>3$ ns).} \label{fig6}
\end{figure}

In Fig. (\ref{fig6}) we see the effects of $U$ on the dynamical
TMR. For $U=1$ meV the TMR is basically the same as before [Fig.
\ref{fig4}(inset)]. For $U=3$ meV the TMR is enhanced (in modulus)
for both on and off voltage regimes ($0<t<3$ ns and $t>3$ ns,
respectively). In particular the Coulomb interaction turns the TMR
even more negative after the bias voltage is turned off, which
reaches $-40$ \% around $3.5$ ns for $U=3$ meV.

\section{Conclusion}

We predict novel spin-dependent effects in a quantum dot coupled to
two ferromagnetic leads driven by a rectangular bias voltage pulse.
Based on nonequilibrium Green function and master equation
techniques we calculated the spin-resolved occupations and currents,
the spin accumulation and the tunnel magnetoresistance in the
transient just after the bias voltage is turned on and off. Our main
findings are: (i) a sign change of the spin accumulation as the time
evolves in the P configuration, (ii) a spike of spin $\uparrow$
current in the emitter lead when the system is antiparallel aligned,
and (iii) a time-dependent TMR that attains negative values. This
negative amount can be further enhanced due to intradot Coulomb
interaction.

The authors acknowledge J. C. Egues and J. M. Elzerman for helpful
discussions. APJ is grateful to the FiDiPro program of the Finnish
Academy for support during the final stages of this work. RMG
acknowledges support from CAPES.


\begin{thebibliography}{99}
\bibitem{wbl} The wideband limit consists of (i) neglecting the real
part of the tunneling self-energy (level-shift), (ii) assuming
that the line-widths  $\G_\s^\eta$ are energy independent
constants, and (iii) allowing a single time dependence, $\D_\eta
(t)$, for the energies in each lead [\onlinecite{hh96}].
\bibitem{commentGamma} In the present work $\G_0 \ll k_B T$, i.e.,
we are at the sequential tunneling limit, so no coherent
oscillations in the current are observed as in previous works
[\onlinecite{apj94}],[\onlinecite{nsw93}].
\end{thebibliography}
\end{document}